\documentclass[preprint,authoryear,12pt]{elsarticle}

\usepackage{graphicx}

\usepackage{amssymb}

\journal{Advances in Space Research}

\begin{document}

\begin{frontmatter}



\title{Broad spectral line and continuum variabilities in QSO spectra
 induced by microlensing of diffusive massive substructure}


\author{Sa\v sa Simi\'c\corref{cor}}
\address{Faculty of Science, University of Kragujevac, Radoja Domanovi\'ca 12, 34000 Kragujevac, Serbia}
\cortext[cor]{Corresponding author}
\ead{ssimic@kg.ac.rs}

\author{Luka \v C. Popovi\'c}
\address{Astronomical Observatory, Volgina 7, 11160 Belgrade, Serbia}
\address{Department of Astronomy, Faculty of Mathematics, University of Belgrade, Studentski trg 16, 11000 Belgrade, Serbia}

\begin{abstract}
We investigate the variability of the continuum and broad lines in  QSO spectra (particularly in the H$\beta$ line and continuum
at $\lambda$ 5100 \AA ) caused by microlensing of a diffuse massive structure (like an open star cluster).
We modeled the continuum and line emitting region and simulate a lensing event by a star cluster located in an intervening galaxy.
Such a type of microlensing event can have a significant influence on magnification and centroid shift of the
broad lines and continuum source. We explore relationships between the continuum and broad line flux variability during
the microlensing event.
\end{abstract}


\end{frontmatter}

\parindent=0.5 cm

\section{Introduction}

The light from a distant quasar (or QSO), can be perturbed by compact massive objects, as e.g.
stellar clusters, intermediate mass compact objects (IMCOs with $10^{2-4} M_\odot$) and cold dark matter
structure, especially in the case of macrolensed QSOs.
This can cause a magnification in the luminosity of a QSO \citep[see e.g.][]{pop05,jov08,sl12,st12} and
in its photometric position \citep[see e.g.][etc.]{Hog95,Walker95,Miyamoto95,Dominik00,Lee10,za10,tr10,popovic13},
the so called lensing effects. Additionally, lensing can affect the spectra of a lensed QSO in the continuum
\citep[see e.g.][]{le98,pop05,bl05,jov08,po12} and broad lines \citep[see e.g.][etc.]{Schneider90,ab02,ri04,sl12,gu13} and
produce the spectral anomaly in the images of a lensed quasar.
Depending on the image angular separation lensing can be divided into several categories from which
three are the most used \citep[see][]{za10,tr10}: macrolensing ($>0.1$ arcsec), millilensing ($\sim 10^{-3}$ arcsec) and
microlensing ($\sim10^{-6}$ arcsec). The effect of lensing of compact objects
(millilensing and microlensing) is usually consider to be present in the images
of a macrolensed QSO, but the effect of micro/millilensing may be present also
in QSOs which are not macrolensed. This is the case where the line-of-sight from an
observer to a source does not lie very close to the center of a massive galaxy, but still the compact
massive object from the galaxy and stars can affect the QSO light \citep[see e.g.][]{zah04,za07}.
Note here that there is a number of QSOs observed close to galaxies, cluster of galaxies and/or throughout a galaxy stellar disc \citep[see e.g.][etc]{ga03,zi07,me10,an13}. Consequently, the spectra of non-(macro)lensed QSOs may be also affected by lensing of compact objects, but here we will consider the case of macro-lensed quasars, taking a typical lens distance as $z_d=0.5$ and source as $z_s=2.0$.

On the other hand, active galactic nuclei (AGNs, QSOs are a branch of AGNs) very
often show an intrinsic variability in the continuum and broad lines that can be used to constrain
the structure of these objects \citep[see e.g.][and reference therein]{pet13}, but variability in the
continuum and broad lines may be caused also by milli/microlensing. Especially if a group of stars acts as
a lens the strong amplification can be seen, not only in the continuum source, but also in broad line source,
so called Broad Line Region -- BLR. Here, similarly to the work of \cite{ga11}, we use the microlensing
event, that produces variation in the continuum and emission from the BLR region, (here we consider the H$\beta$ broad
line) to study the properties of the BLR region.

The aim of this paper is to investigate correlation between the  variability of broad lines and continuum during
lensing by a bulk of stars concentrated on a relatively small surface, as e.g. star clusters, simulating a
massive diffuse lensing structure that contains between several tens to several hundreds of solar mass stars.
We consider the complex emitting structure of a QSO, taking stratification in the continuum and the BLR emission.
We explore the microlensing influence on the $H\beta$ spectral line emitted from the BLR
and optical continuum at $\lambda$ 5100\AA.

The paper is organized as follows: In the next section we present the  source and lens model; in \S 3 we give
results of our simulations and discussion, and finally in \S 4 we outline our conclusions.

\section{Source and lens models}
\label{sec:srcmodel}

\subsection{The continuum emission}

An AGN has a complex inner structure, meaning that different parts are emitting in different spectral bands,
i.e. a wavelength dependent dimension of different emission regions is present in AGNs \citep[see e.g.][]{pop12}.
It is widely accepted that the majority of the AGN  radiation is coming from the accelerated material spiraling down towards
a black hole in a form of an accretion disc. The radiation from the disc is mostly thermalized from outer regions
$R_{out}$ to the center, with slight exception at the inner radius, i.e. very close to the black hole, where Compton
upscattering due to the high mass accretion could have a significant role \citep[see][]{Done11}.

Here we use the same model of an accretion disc as in \cite{popovic13} for the continuum, taking
into account the spectral stratification
of the source where the effective temperature is a function of the radius \citep[see][]{Krolik98,popovic13}:

$$T\propto R^{-3/4}(1-(R_{in}/R)^{1/2})^{1/4},$$
where $R_{in}$ is the inner disc radius. At a larger radius this equation could be reduced to $T\propto R^{-\beta}$,
where in the standard model $\beta = 3/4$.

The model allows us to calculate the luminosity of a small surface element at an arbitrary position in the
disc. It is proportional to the surface
energy density and area of the emitting surface \citep{popovic13}:

\begin{equation}
\label{eqn:source}
dL(\lambda, R)\propto \frac{dS}{\lambda^{5}} (exp(\frac{hc}{\lambda k \alpha(\beta) R^{-\beta}}-1))^{-1}.
\end{equation}
where $dS$ is the surface element of the source and $h$ and $k$ are the Planck and Boltzman constant, respectively.
We replaced $T$ in the expression for the energy density
with the distance $R$ and computed the proportionality coefficient $\alpha(\beta)=T_0 R_0^{\beta}$,
where $T_0$ is the temperature at the distance $R_0$. To compute the spectral energy distribution
(SED) for disc configuration we integrate over the whole disc area:

\begin{equation}
\label{eqn:source_int}
L(\lambda)\propto \int_{S_{disc}}^{}dL(\lambda,R)
\end{equation}

The inclination of the disc with respect to the observer could be included as $cos(i)$
(\emph{i}-inclination angle). In this paper we assumed a face-on disc ($i=0$).

Using the disc model for the UV and optical continuum  emission of QSOs we can explore influence of the
microlensing on such system and model of SED amplification  and fluctuation in object position as it has been shown
in \cite{popovic13}. But, here, especially we will pay attention
only to continuum around the H$\beta$ line.

\begin{figure}
\centering
\includegraphics[width=4cm]{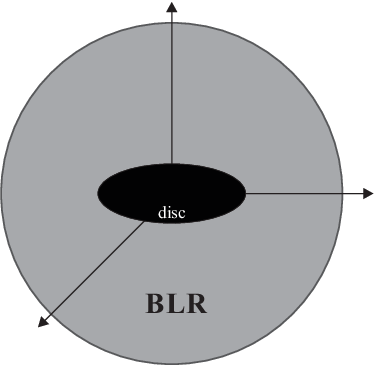}
\includegraphics[width=4cm]{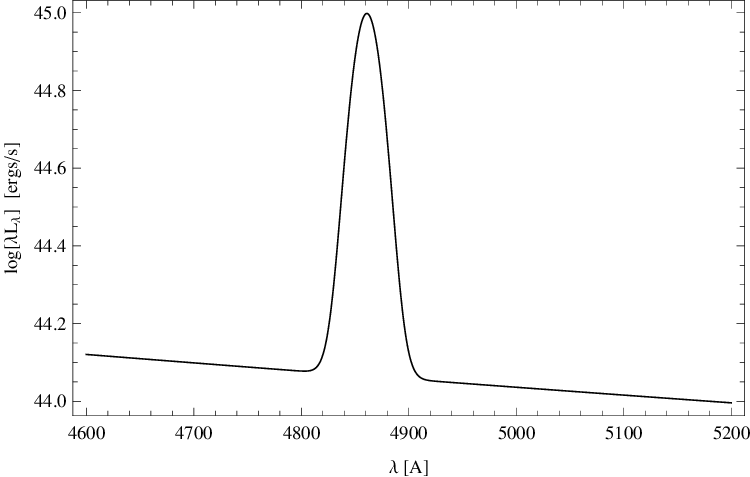}
\caption{First panel: The model of the source, disc-continuum and the BLR region. Second panel:
The composite spectra in the H$\beta$ wavelength range.}
\label{fig_source}
\end{figure}

\subsection{The model of the BLR and H$\beta$ line}
\label{sec:blrmodel}

The lensing effect is geometrical, and its influence on spectra is caused by different sizes of emission regions
\citep{pop05}. To explore qualitatively relationships between the broad line and continuum flux variation we choose
the H$\beta$ wavelength range, since
one can expect that this effect will be similar in other broad lines and the corresponding continuum.

We accepted that the emission gas
in the BLR is virialized and that photoionization is dominant in the BLR, consequently,
the BLR size and the line properties depend on the mass of the central object \citep{pw99}.

To add the BLR emission to the above described continuum disc-source, we assumed a model of
spherically distributed clouds, taking that some constraints for the BLR are connected with the
luminosity of the continuum source. Kinematical parameters of the BLR  directly depends on the dimension
and mass of the central black hole.

The relationship between the BLR size (R$_{BLR}$) and continuum luminosity has been taken from \cite{ka05}:
\begin{equation}
\frac{R_{BLR}}{10 {\rm \ lt\,days}}= (2.23\pm0.21)
\left(\frac{\lambda L_{\lambda}(5100\,{\mbox{\AA}} )}{\rm 10^{44} \ ergs\,s^{-1}} \right)^{0.69\pm0.05}.
\label{eq:blr_size}
\end{equation}
where $L_{\lambda}(5100\,{\mbox{\AA}} )$ is the continuum luminosity at $\lambda$ 5100\AA.

To estimate the intensity of the H$\beta$ line, we used relation \citep[][]{ka05}:
\begin{equation}
\frac{R_{BLR}}{10 {\rm \ lt\,days}} = A L^{B}
\label{eq:Hb_lum}
\end{equation}
where $A=8.91^{+0.92}_{-0.83}$ and  $B=0.690\pm0.068$ are constants taken from \cite{ka05}.
Estimating  the BLR size from Eq.
\ref{eq:blr_size}, we can calculate the luminosity of the H$\beta$ line as:

\begin{equation}
\label{eq:LHbeta}
\log{L(H\beta)\over{10^{43}}}= 1.45\log\{{{R_{BLR}\over{10\ lt\,days}}-8.91}\}.
\end{equation}

In order to estimate the equivalent width (EW) of the H$\beta$ we assumed that the broad emission
lines are broadened primarily by
the virial gas motions in the gravitational potential of the central black hole \citep[see][]{pw99}.
Therefore, we assumed that the dimension of the BLR is from the continuum source to the above estimated $R_{BLR}$
and that in this volume are located N uniformly distributed clouds with velocities $v_i$ which
depend from the mass of the black hole and distances from the center as:

\begin{equation}
\label{eq:vi}
v_i=\sqrt{2GM_{BH}\over{R_i}}.
\end{equation}
where $v_i$ is taken as Gaussian dispersion velocity (since the line profile is assumed to have a Gaussian profile).
The width of the $H_\beta$ line  is assumed as 2000 km/s.

Here we assumed that the BLR is homogenous  and to represent the width of  the $H_{\beta}$ line we use velocity
dispersion of a Gaussian as at distance of the averaged radius of the BLR. The $H_{\beta}$ intensity is taken to be
directly proportional to the volume of the BLR  $L_{H\beta} \sim V_{BLR},$. Note here that the model of the BLR is
very simple, but since we explore the flux variability (amplification) that depends on the BLR and continuum source dimensions,
it should not significantly affect obtained results.

\subsection{Microlensing model}

\begin{figure}
\centering
\includegraphics[width=13.5cm]{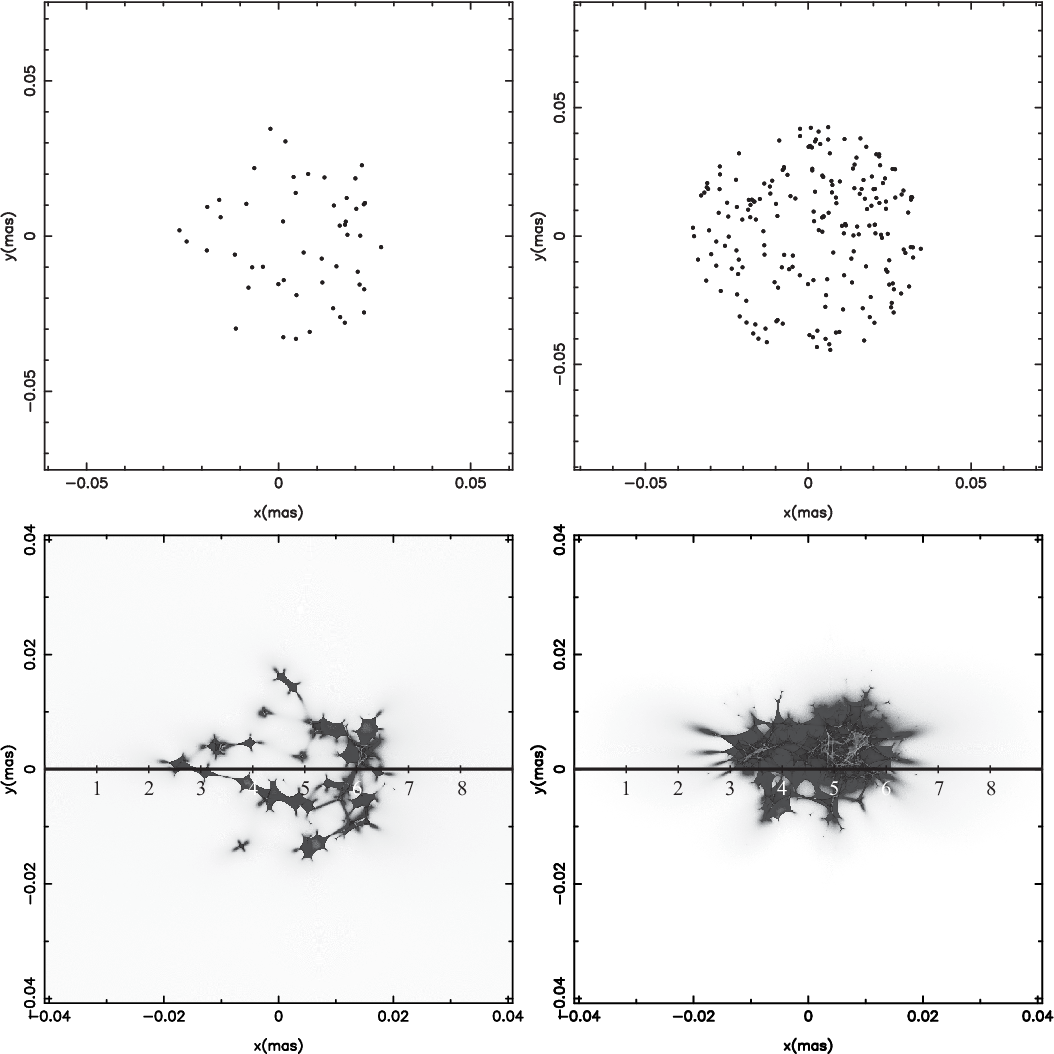}
\caption{Upper panels show a randomly star distribution in the lens plane, a) case 50 solar mass stars distributed randomly
in a circle with radios of 0.04 mas in lens plane (left panels) and b) the same as a) but for
200 solar mass stars (right panels). In the bottom panels we give the corresponding magnification maps in the source plane.
The solid lines across the microlens map show the path of the source, numbers denote
the position for which spectra are shown in Fig. \ref{fig:dfc_lum_snum0}. The standard lens system
with $z_d=0.5$ and $z_s=2.0$ is considered.}
\label{fig:stars_map}
\end{figure}

A distribution of stars in the lens plane is used to generate microlensing magnification map in the
source plane which is computed by the ray-shooting technique \citep[see e.g.][]{Kayser86,Schneider86,Schneider87,Treyer04}.
This technique is in details described in \cite{popovic13} and here will not be repeated. In Fig. \ref{fig:stars_map}
we present star distributions (panels up) for 50 and 200 solar mass stars, and corresponding
amplification maps (panels down). The number and random distribution of stars in our model is typical for
open star clusters. Note here that globular star clusters are more compact in size and much more populated than the open clusters.
Consequently this produce microlensing effect to be very similar as the point like objects, with well studied microlensing influence.

\begin{figure*}
\centering
\includegraphics[width=13.5cm]{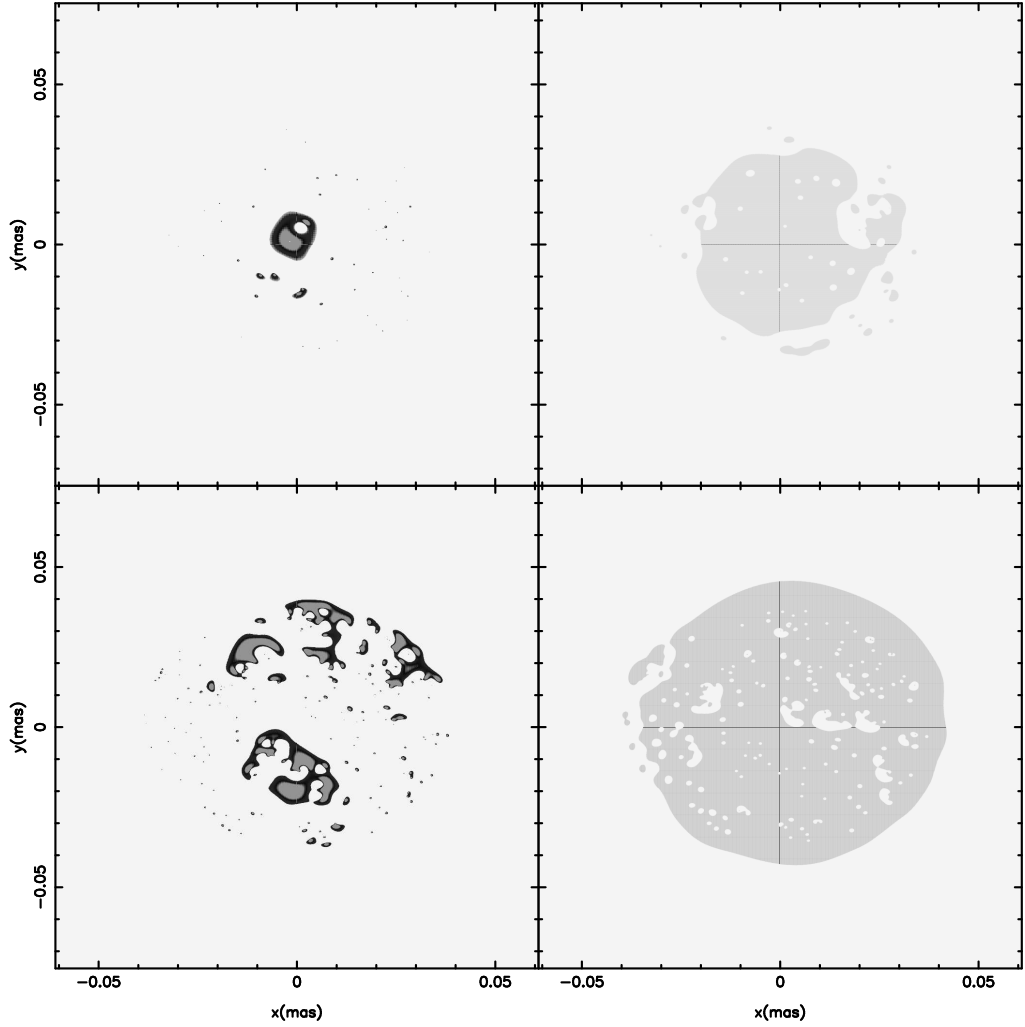}
\caption{Images of the continuum (left) source and the BLR (right).
The lens maps are presented in Figure \ref{fig:stars_map}. Panels up present the case of
50 solar mass star lensing, while panels down present 200 solar mass lensing. The lens system is assumed to have
$z_d=0.5$ and $z_s=2.0$.}
\label{fig:map_gl}
\end{figure*}

Using the described model for source and microlens we are able to create images of the lensed continuum disc-source
and BLR (see Fig. \ref{fig:map_gl}). Consequently,
we are able to calculate the centroid shift of the image for different spectral filters as \cite[][]{popovic13}:

\begin{equation}
\label{eqn:cent_shift}
D_{cs}(F)=\frac{\int_{F}\sum_{npix}x_{pix}L_{pix}^{lens}(\lambda) d\lambda}{\int_{F}\sum_{npix}L_{pix}^{lens}(\lambda)d\lambda}
\end{equation}
where $F$ denotes integration for a particular (continuum and broad line) spectrum and A is the whole energy range. Also,
we designated with $x_{pix}$ the coordinate of a particular pixel in the image with the corresponding luminosity of $L_{pix}^{lens}$.

The magnification for a particular source image is computed as the ratio of the  luminosity for all pixels in a spectrum
with the luminosity in the same spectrum without lens influence, as \cite[][]{popovic13}:

\begin{equation}
\label{eqn:UBVR_magnification}
m(F) = \frac{\int_{F}\sum_{npix}L_{pix}^{lens}(F)d\lambda}{\int_{F}\sum_{npix}L_{pix}^{nolens}(F)d\lambda}.
\end{equation}

With these variables computed for any particular image during the lensing event (transition) we are able to estimate its
maximal and minimal values, as well as the trend of change for different lens condition.

The relevant length scales for microlensing is the dimension of the Einstein Ring Radius (ERR) in the lens plane, defined as:
\begin{equation}
\label{eqn:xi0}
\xi_0=\sqrt{\frac{4Gm}{c^2}\frac{D_d D_{ds}}{D_s}},
\end{equation}
and its projection in the source plane is:
\begin{equation}
\label{eqn:err}
ERR=\frac{D_s}{D_d}\xi_0=\sqrt{\frac{4Gm}{c^2}\frac{D_{s}D_{ds}}{D_d}}
\end{equation}
where $G$ is the gravitational constant, $c$ is the speed of light, $m$ is the microlens mass.
We adopted standard notation for cosmological distances to the lens $D_d$, source $D_s$ and between them $D_{ds}$.
In our simulation value of 1ERR is close to $5\cdot 10^{16}$ cm $\sim 1.7$ pc $\sim 20.6$ lt-days.

Time scales for a microlensing event is described in detail in the book on gravitational microlensing
\citep[see][]{sch92,Zah97,Pett01}. Based on the sizes of the source ($R_{src}$) and caustic ($r_{caustic}$)
pattern we distinguish two cases, when $R_{src}>r_{caustic}$ and $R_{src}<r_{caustic}$ (presented in detail in \cite{jov08}).
Both cases could be expressed in a single form:

\begin{equation}
\label{eqn:time_scale}
t_{crossing}=(1+z_d)\frac{R}{v_{\perp}(D_s/D_d)}
\end{equation}
where $R$ replaces the $R_{src}$ or $r_{caustic}$. Here we used a simple approach,
since we considered that a microlensing event duration corresponds to the time needed for crossing over
the caustic network created by the lens. In this way the dimension of the caustic patterns in the magnification map
determines the total time scale for a particular event, and it can be computed by using  Eq. \ref{eqn:time_scale},
with the $R$ replaced by the dimension of map $r_m$. We used already introduced comoving distances $D_s$ and given
map dimensions in ERRs to calculate map linear dimensions, and hence the width of the caustic network.
$v_{\perp}$ is transverse velocity, and in our simulations we assumed the typical one of  $v_{\perp}=600.0$ km/s,
that gives a crossing time for complete map (40 ERR) of the order of thousands years.

\subsection{Parameters of source and lens}

The continuum source is defined with it's inner and outer radius. As we adopted the standard
model for the disc we considered that the most of the continuum radiation in the observed energy range,
is coming from the disc part within the area defined by $R_{in}=10^{13}$m and $R_{out}=10^{15}$m.
For evaluating the proportionality coefficient $\alpha$ (see Eq. \ref{eqn:source}),
we took the temperature of $T_0=2\cdot10^4$K, with the peak radiation around 1500\AA,
at the radius $R_0=3.15\cdot10^{13}$m, \citep[see][]{Blackburne11}. The coefficient $\beta$ usually has the value
of $\beta=3/4$ and we kept that value constant throughout all simulations. The disc
inclination can be changed, but in the  simulations we assumed a face-on disc orientation.
The source plane dimension is the same in all our computation and equal to 40 ERR, that
with the size of the map of 2000 pixels gives a resolution of 0.02 ERR/pix.

By evaluating the continuum source luminosity at $\lambda=$5100 \AA\  we
are able to estimate the dimensions of BLR region using the
Eq. \ref{eq:blr_size} and hence line parameters Eqs. \ref{eq:LHbeta} and \ref{eq:vi}.

Here we consider the BLR size of 128 light days (i.e. the radius is 64 light
days), for a black hole of $8\cdot 10^8 \ M_\odot$. The luminosity of the H$\beta$ line without lensing is  $1.5\cdot 10^{44}$ ergs/s.

The lens has been assumed to have a circular shape, containing $N_{s}$ stars of Solar mass
ranging from 50 and 200 solar mass stars. We also assumed that lens and source are placed
at the cosmological distances, with $z_d=0.5$ and $z_s=2.0$ (so called the standard lens).
We are confident, based on the discussion in \cite{popovic13} that such carefully chosen lens reflect
good enough condition for the gravitational bound systems. Any more massive and densely populated lens
will act as one compact object with known influence on the distant source.

In all calculations we assumed a flat cosmological model, with $\Omega_{M}=0.27$, $\Omega_{\Lambda}=0.73$
and $H_{0}= 71\ \rm km\ s^{-1} Mpc^{-1}$.

\section{Results}

To explore correlation between the line and continuum variability during microlensing event by diffuse massive structure
we modeled microlensing of the described above source by a group of 50 and 200 solar mass stars. As it can be seen in Fig.
\ref{fig:map_gl}, the images of the continuum source and BLR are quite different. In the case of 50 stars there are microlensing magnification
of both regions, and magnifications stay significant in the more massive lens (200 stars).
It is interesting that the continuum source is splited in several spots, but the BLR image shows several holes.
As it can be seen in Fig. \ref{fig:map_gl}, in the surface of a source (in the BLR and continuum), at
some parts, the demagnification is present. This demagnification is caused by the distribution of caustics in the lens,
but the total brightness of the source is always magnified.

\begin{figure}
\centering
\includegraphics[width=13.5cm]{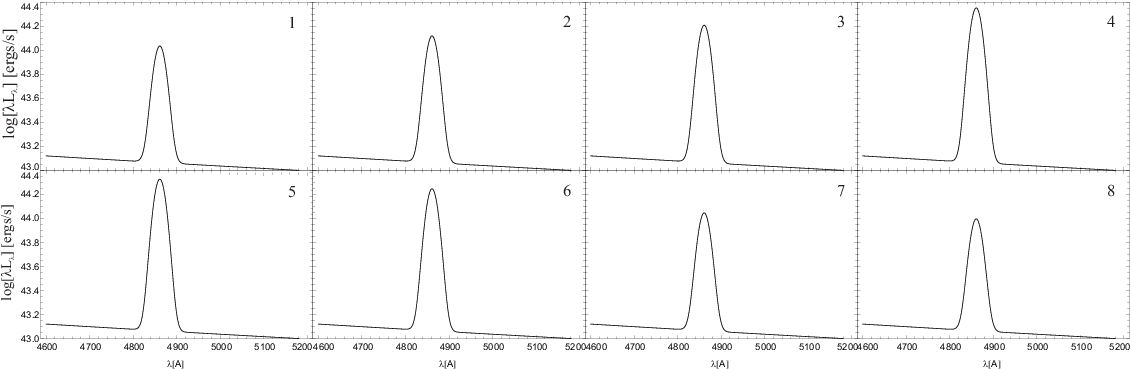}
\includegraphics[width=13.5cm]{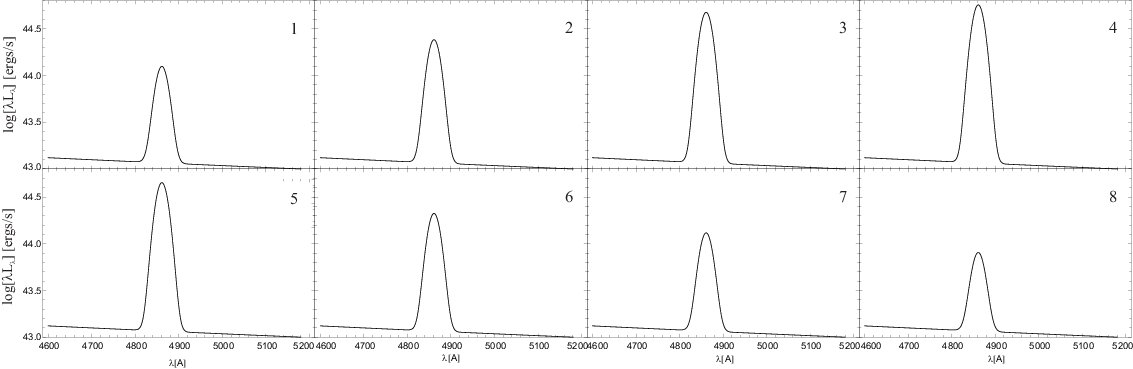}
\caption{The variation in the H$\beta$ spectrum range during a microlensing event: the case of 50 (panel up) and 200 (panel down)
solar mass stars. The number on plots corresponds to the position shown in Fig. \ref{fig:stars_map}.}
\label{fig:dfc_lum_snum0}
\end{figure}

We also explore a transition of the lens along such a complex sources and changing in spectral properties.
In Fig. \ref{fig:dfc_lum_snum0} we present the spectral changes due to microlensing the systems of 50 (panels up) and
200 (panels down) solar like stars. The particular image in the panels refers to the positions denoted with
numbers in Figure \ref{fig:stars_map}, which present points of our calculation. A rough estimate gives that these
points cover the time interval of around 1500 days, starting from left position and going toward the edge on right side.
As it can be seen in Fig. \ref{fig:dfc_lum_snum0}, very important changes are seen in the line amplification,
while there is no significant changes in the line profile.

\subsection{Continuum vs. line variation}

\begin{figure}
\centering
\includegraphics[width=6.5cm]{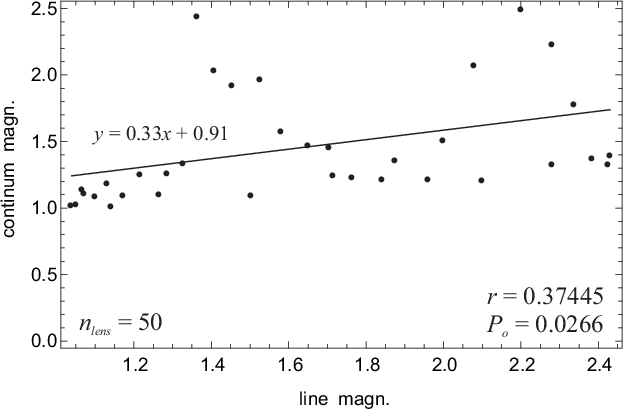}
\includegraphics[width=6.5cm]{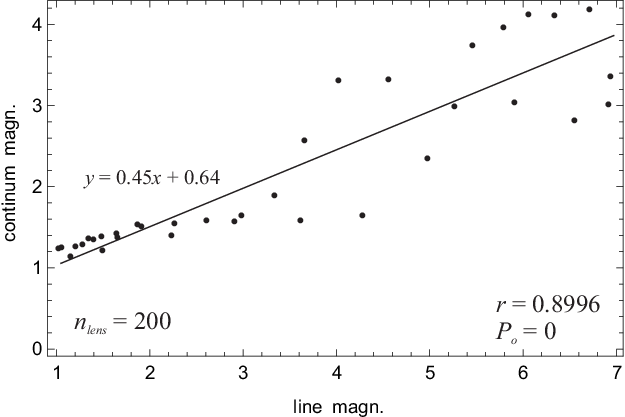}
\caption{Magnification in line against magnification in the continuum for the lens of 50 (left) and 200 (right) solar
mass stars.}
\label{cor}
\end{figure}

As expected during the lensing event, a variations (amplification) in the broad line and continuum are present.
Those variations can be used to explore the parameters of the BLR \citep[see
e.g.][]{ga11}. In our case
we consider a very simple BLR model, taking that the BLR has one velocity field over entire volume.
The correlations between the continuum and line
luminosities for considered two cases are given in Fig.
\ref{cor}, a weak correlation is present in the case of 50 star lens ($r=$0.37445,
P=0.0266), that is not statistically important (left panel in Fig.
\ref{cor}), while in 200 star lens ($r=$0.8996, P=0) the correlation is
important and statistically significant.

\begin{figure}
\centering
\includegraphics[width=6.5cm]{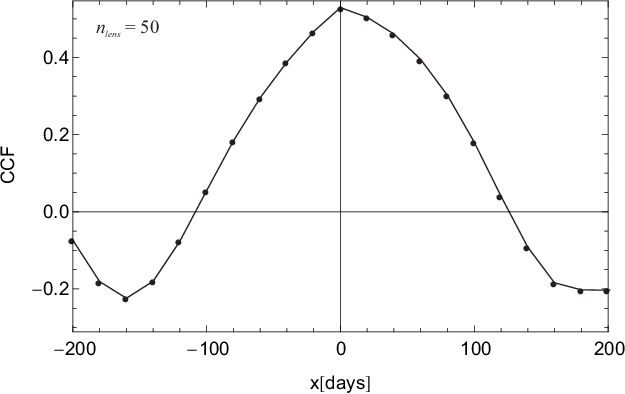}
\includegraphics[width=6.5cm]{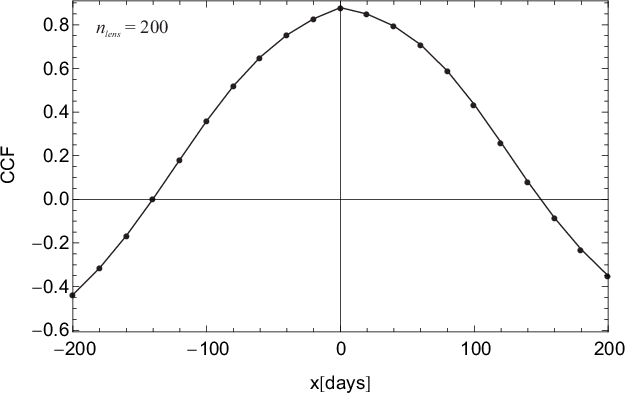}
\caption{CCFs between the continuum and line for the case of lensing of 50 (left) and 200 (right) solar
mass stars.  The X-axis presents
the distance scale in the source plane. }
\label{ccf}
\end{figure}

On the other hand one can expect that, during a microlensing event, the
delay of the signal between the continuum and line could be observed. Therefore
we calculated 40 points across the paths shown in Fig.
\ref{fig:stars_map}, and calculated cross correlation functions (CCF) for both cases
(lensing 50 and 200 solar mass stars, Fig. \ref{ccf} -- left and right panel,
respectively). As one can see from Fig. \ref{ccf}, the lags are zero, i.e.
there is no lag that indicates the size of the BLR (64 light days).

The reason for it is that the continuum source stays highly magnified during
the event, and variation in the continuum is relatively small with
regards to variation in the line. One can expect that a smaller source, as it is
the continuum should have a more prominent variation during a typical microlensing event
\citep[several stars, see e.g.][]{jov08}, but in this case the continuum stays high amplified during the event,
since there is always a group of caustics that crosses the small continuum source. However, in the case of lens with 50 stars, one can see in Fig. \ref{cor} that the amplification in the continuum can be significantly higher than in the line
(points that have a large scatter in Fig. \ref{cor} left). The amplification in the line is higher than in the continuum, that is expected,
since the continuum source is smaller, and always covered by a smaller group of caustics than line one. As e.g.
in the lens with 200 stars, the density of caustics is higher in the center, but a small surface of the continuum source will be amplified
by smaller number of caustics than the BLR.
The ERR of the lens is comparable with the projected dimension of the BLR, therefore, amplification in the
line flux shows higher variability during lensing than the continuum one.

\begin{figure}
\centering
\includegraphics[width=10cm]{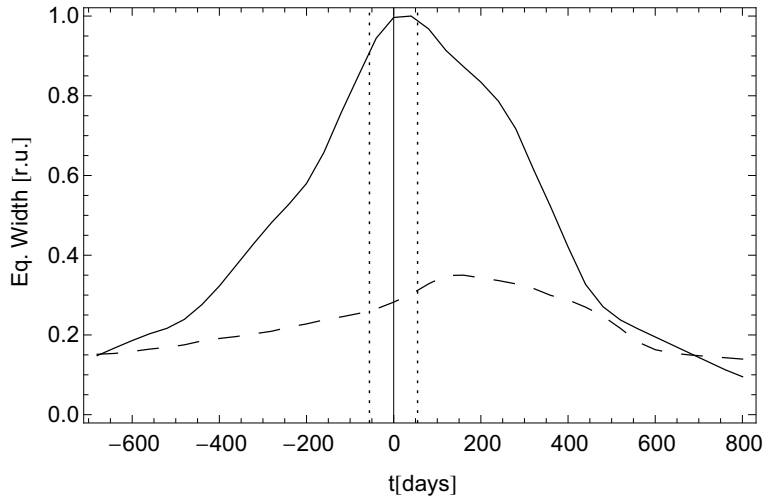}
\caption{Equivalent width of the H$\beta$ line as a function of distance
scale in the source plane (given in light days). Full line -
case with 200 stars and dashed line case with 50 stars.
Verticale dashed lines denote the
edge of the BLR - 128 l.d. (64 l.d. left and right from the center)}
\label{fig:dfc_lum_snum}
\end{figure}

As it can be seen in Fig. \ref{fig:dfc_lum_snum} the equivalent width
is increasing during the event, and this increase is lasting after the
center of lens crosses the center of source, and after that decreases  (see Fig. \ref{cor}).
As we noted above, it
is since the continuum source is too compact, and all time stays
amplified, but the BLR has maximal amplification when the lens crossing
near the central part.

As one can see in Fig. \ref{fig:dfc_lum_snum}, the EW, in both cases, has an
off-centered maximum, that is close to  the BLR radius (vertical
dashed line in Fig.  \ref{fig:dfc_lum_snum}). This is caused by the fact
that crossing the source (continuum + BLR), amplifications of the BLR and
continuum increase, but when the continuum source is completely covered by the
lens and the amplification of the continuum stays more or less constant, the
lens covering rest part of the BLR and amplification in the line is
increasing (after the lens crossed the central part of the source). Therefore, the peak of the EW curve and
its asymmetry may indicate the BLR sizes.

At the end we should note here that the intrinsic variation of an AGN can be present and can affect the obtained relationships between the broad line and continuum variability in the case of microlensing.

\section{Conclusion}

We modeled lensing of a complex source (typical AGN) containing the continuum source suppose to be a disc and BLR assumed to
be spherically symmetric with a homogenous velocity field. The lens is assumed to be a massive diffuse structure (like a stellar
cluster) containing 50 and 200 solar mass stars. Our simulation were performed for a standard lens system
with $z_d=0.5$ and $z_s=2.0$. We measured the variability in the broad line and corresponding continuum, in this case of
the  broad H$\beta$ line and continuum at $\lambda$5100\AA.

From our simulations we can outline following conclusions:

i) The diffuse massive lens can significantly magnify, in addition to the continuum, the broad line emission. Also, the
images in the continuum and in the broad line are different in shape.

ii) The amount of variability in the broad line flux is higher than in the continuum,
since the continuum source is much compact and stays strongly
magnified during a longer period of lensing.

iii) The correlation between the line and continuum luminosity variations is higher in the case of more massive diffuse object, e.g.
in the case of lensing of 50 solar mass stars the coefficient of correlation $r=$0.37445  (P=0.0266),
and for 200 stars is significantly higher $r=$0.8996 (P=0) and statistically significant.

iv) CCFs of the continuum with line luminosity during microlensing event show that the measured lag
does not  correspond to the BLR dimensions, in our case we obtained zero lag; but it seems that the
EW curve during the event could give more information about the BLR sizes,
but this should be explored on a larger number of different models for the
lens as well as for the BLR.

\section*{Acknowledgments}

This work is a part of the project (176001) "Astrophysical Spectroscopy of Extragalactic Objects,"
supported by the Ministry of Science and Technological Development of Serbia.



\begin{thebibliography}{}

\bibitem[Abajas et al.(2002)]{ab02} Abajas, C., Mediavilla, E., Mu\~noz, J. A., Popovi\'c, L. \v C.., Oscoz, A.,The Influence of Gravitational Microlensing on the Broad Emission Lines of Quasars, ApJ, 576, 640-652, 2002.

\bibitem[Andrews et al.(2013)]{an13}
Andrews, H., Barrientos, L. F., L\'opez, S., Lira, P., Padilla, N., Gilbank, D. G., Lacerna, I., Maureira, M. J., Ellingson, E.,
Gladders, M. D., Yee, H. K. C., 	
	Galaxy Clusters in the Line of Sight to Background Quasars. III. Multi-object Spectroscopy,
	ApJ, 774,  article id. 40, 35 pp., 2013.

\bibitem[Blackburne et al.(2006)]{bl05}
Blackburne, J. A., Pooley, D., Rappaport, S.,
X-Ray and Optical Flux Anomalies in the Quadruply Lensed QSO 1RXS J1131-1231, ApJ, 640, 569-573, 2006.

\bibitem[Blackburne et al.(2011)]{Blackburne11} Blackburne, J. A.,  Pooley, D.,  Rappaport,  S. \& Schechter, P. L., Sizes and Temperature Profiles of Quasar Accretion Disks from Chromatic Microlensing, ApJ, 729, 34-54, 2011.

\bibitem[Done et al.(2012)]{Done11} Done C., Davis, S.W., Jin, C, Blaes,
O. \& Ward, M., Intrinsic disc emission and the soft X-ray excess in active galactic nuclei, MNRAS, 420, 1848-1860, 2012.

\bibitem[Dominik \& Sahu(2000)]{Dominik00} Dominik, M. \& Sahu, K., Astrometric Microlensing of Stars, ApJ, 534, 213-226, 2000.

\bibitem[Gazta\~naga(2003)]{ga03}
Gazta\~naga, E., Correlation between Galaxies and Quasi-stellar Objects in the Sloan Digital Sky Survey:
A Signal from Gravitational Lensing Magnification?, ApJ, 589,  82-99, 2003.
	
\bibitem[Garsden et al.(2011)]{ga11} Garsden, H., Bate N. F. and Lewis, G. F., Gravitational Microlensing of a Reverberating Quasar
Broad Line Region – I. Method and Qualitative Results, MNRAS, 418, 1012-1027, 2011.

\bibitem[Guerras et al.(2013)]{gu13} Guerras, E., Mediavilla, E., Jimenez-Vicente, J., Kochanek, C. S., Mu\~noz, J. A., Falco, E., Motta, V.,
Microlensing of Quasar Broad Emission Lines: Constraints on Broad Line Region Size, ApJ, 764, 160-169, 2013.

\bibitem[Hog et al.(1995)]{Hog95} Hog, E., Novikov, I. D. \& Polnarev, A. G., MACHO photometry and astrometry, A\&A, 294, 287-294, 1995.

\bibitem[Jovanovi\'c  et al.(2008)]{jov08} Jovanovi\'c, P., Zakharov, A. F., Popovi\'c, L. \v C., Petrovi\'c, T., Microlensing of the X-ray, UV and optical emission regions of quasars: Simulations of the time-scales and amplitude variations of microlensing events, MNRAS, 386, 397-408, 2008.

\bibitem[Kaspi et al.(2005)]{ka05}Kaspi, S., Maoz, D., Netzer, H., Peterson, B. M., Vestergaard, M., Jannuzi, B. T., The Relationship between Luminosity and Broad-Line Region Size in Active Galactic Nuclei, ApJ, 629, 61-71, 2005.

\bibitem[Kayser et al.(1986)]{Kayser86} Kayser, R., Refsdal, S. \& Stabell, R., Astrophysical applications of gravitational micro-lensing, A\&A, 166, 36-52, 1986.

\bibitem[Krolik(1998)]{Krolik98} Krolik, J. H., Active Galactic Nuclei: From the Central Black Hole to the Galactic Environment (Princeton University Press), 1998.

\bibitem[Lee et al.(2010)]{Lee10} Lee, C. H., Seitz, S., Riffeser, A. \& Bender, R., Finite-source and finite-lens effects in astrometric microlensing, MNRAS, 407, 1597-1608, 2010.

\bibitem[Lewis et al.(1998)]{le98} 	Lewis, G. F., Irwin, M. J., Hewett, P. C., Foltz, C. B., Microlensing-induced spectral variability in Q 2237+0305, MNRAS, 295, 573-586, 1998.

\bibitem[Meusinger et al.(2010)]{me10} Meusinger, H., Henze, M., Birkle, K., Pietsch, W., et al. J004457+4123 (Sharov 21): not a remarkable nova in M 31 but a background quasar with a spectacular UV flare, A\&A, 512A, 1-15, 2010.

\bibitem[Miyamoto \& Yoshii(1995)]{Miyamoto95} Miyamoto, M. \& Yoshii Y., Astrometry for Determining the MACHO Mass and Trajectory, AJ, 110, 1427-1433, 1995.

\bibitem[Paczynski(1986)]{p86}
	Paczynski, B.,
	Gravitational microlensing by the galactic halo, ApJ, 304, 1-5, 1986

\bibitem[Peterson \& Wandel(1999)]{pw99}Peterson, B.M. \& Wandel, A., Keplerian Motion of Broad-Line Region Gas as Evidence for Supermassive Black Holes in Active Galactic Nuclei, ApJL, 521, 95-98, 1999.

\bibitem[Peterson(2013)]{pet13}Peterson, B.M., Measuring the Masses of Supermassive Black Holes, Space Science Reviews, on line, DOI:10.1007/s11214-013-9987-4.

\bibitem[\protect\citeauthoryear{Petters et al}{2001}]{Pett01} Petters A. O., Levine H. \& Wambsganss J., Singular Theory and Gravitational
Lensing, Birkhauser, Boston, 2001.

\bibitem[Poindexter et al.(2008)]{Poindexter08} Poindexter, S., Morgan, N. \& Kochanek, C. S., The Spatial Structure of an Accretion Disk, ApJ, 673, 34-38, 2008.

\bibitem[Popovi\'c \& Chartas(2005)]{pop05} Popovi\'c, L. \v C., Chartas, G., The influence of gravitational lensing on the spectra of lensed QSOs, MNRAS, 357, 135-144, 2005.

\bibitem[Popovi\'c et al.(2012)]{pop12} Popovi\'c, L. \v C., Jovanovi\'c, P., Stalevski, M., Anton, S., Andrei, A. H., Kova\v cevi\'c, J., Baes, M., Photocentric variability of quasars caused by variations in their inner structure: consequences for Gaia measurements, A\&A, 538A, 107-117, 2012.

\bibitem[Popovi\'c \& Simi\'c(2013)]{popovic13}Popovi\'c, L.\v C. \& Simi\'c, S., Spectro-photometric variability of quasars caused by lensing of diffuse  massive substructure: Consequences on flux anomaly and precise astrometric measurements, MNRAS, 432, 848-856, 2013.

\bibitem[Pooley et al.(2012)]{po12} Pooley, D., Rappaport, S., Blackburne, J. A., Schechter, P. L., Wambsganss, J., X-Ray And Optical Flux Ratio Anomalies In Quadruply Lensed Quasars. II. Mapping the Dark Matter Content in Elliptical Galaxies, ApJ, 744, 111-124, 2012.

\bibitem[Richards  et al.(2004)]{ri04} Richards, G. T., Keeton, C. R., Pindor, B. et al., Microlensing of the Broad Emission Line Region in the Quadruple Lens SDSS J1004+4112, ApJ, 610, 679-685, 2004.

\bibitem[Sluse et al.(2012)]{sl12} Sluse, D., Hutsem\'ekers, D., Courbin, F., Meylan, G., Wambsganss, J., Microlensing of the broad line region in 17 lensed quasars, A\&A, 544A, 62-89, 2012.

\bibitem[Schneider \& Weiss(1986)]{Schneider86} Schneider, P. \& Weiss, A., The two-point-mass lens - Detailed investigation of a special asymmetric gravitational lens, A\&A, 164, 237-259, 1986.

\bibitem[Schneider \& Weiss(1987)]{Schneider87} Schneider, P. \& Weiss, A., A gravitational lens origin for AGN-variability? Consequences of micro-lensing, A\&A, 171, 49-65, 1987.

\bibitem[Schneider \& Wambsganss(1990)]{Schneider90} Schneider, P. \& Wambsganss, J., Are the broad emission lines of quasars affected by gravitational microlensing?, A\&A, 237, 42-53, 1990.

\bibitem[\protect\citeauthoryear{Schneider et al}{1992}]{sch92} Schneider, P., Ehlers, J., \& Falco, E.E., Gravitational Lenses,
Springer-Verlag Berlin -- Heidelberg -- New York, 1992.

\bibitem[Stalevski et al.(2012)]{st12} Stalevski, M., Jovanovi\'c, P., Popovi\'c, L. \v C., Baes, M., Gravitational microlensing of AGN dusty tori, MNRAS, 425, 1576-1584, 2012.

\bibitem[Treyer \& Wambsganss(2004)]{Treyer04} Treyer, M. \& Wambsganss, J., Astrometric microlensing of quasars. Dependence on surface mass density and external shear, A\&A, 416, 19-34, 2004.

\bibitem[Treu(2010)]{tr10} Treu, T., Strong Lensing by Galaxies, ARA\&A, 48, 87-125, 2010.

\bibitem[Walker(1995)]{Walker95} Walker, M. A., Microlensed Image Motions, ApJ, 453, 37-40, 1995.

\bibitem[Zackrisson \& Riehm(2007)]{za07}Zackrisson, E., Riehm, T. High-redshift microlensing and the spatial distribution of dark matter in the form of MACHOs, A\&A, 475, 453-465,  2007.

\bibitem[Zackrisson \& Riehm(2010)]{za10} Zackrisson, E., Riehm, T., Gravitational Lensing as a Probe of Cold Dark Matter Subhalos, AdAst, ID 478910, 2010.

\bibitem[\protect\citeauthoryear{Zakharov}{1997}]{Zah97}Zakharov A. F., Gravitational Lenses and Microlensing, Janus-K, Moscow, 1997.

\bibitem[Zakharov et al(2004)]{zah04} Zakharov, A.F., Popovi\'c, L. \v C., Jovanovi\'c, P., On the contribution of microlensing to X-ray variability of high-redshifted QSOs, A\&A, 420, 881-888, 2004.

\bibitem[Zibetti et al(2007)]{zi07}	
	Zibetti, S., M{\'e}nard, B., Nestor, D. B., Quider, A. M., Rao, S. M., Turnshek, D. A., 	
	Optical Properties and Spatial Distribution of Mg II Absorbers from SDSS Image Stacking, ApJ, 658,  161-184, 2007.
	








\end{thebibliography}
\end{document}